\documentclass[aps, ams,nofootinbib,twocolumn,superscriptaddress,showpacs]{revtex4}
\usepackage{graphicx}
\usepackage{amsfonts}
\usepackage{psfrag}
\usepackage{amsmath}
\usepackage{float}
\usepackage{amssymb}  
\newcommand{\va}{\scriptscriptstyle}

\def\be{\begin{equation}}
\def\ee{\end{equation}}
\def\ba{\begin{eqnarray}}
\def\ea{\end{eqnarray}}

\def\ut#1{\rlap{\lower1ex\hbox{$\sim$}}#1{}}

\DeclareFontFamily{U}{rsfs}{}         
\DeclareFontShape{U}{rsfs}{m}{n}{<5> rsfs5 <6><7> rsfs7          %
  <8><9><10><10.95><12><14.4><17.28><20.74><24.88> rsfs10}{}     %
\DeclareMathAlphabet{\mathfs}{U}{rsfs}{m}{n}                     %
\newcommand{\mfs}[1]{\mathfs {#1}}                               %
\newcommand{\n}{{\nonumber}}

\newcommand{\sN}{{\mfs N}}

\newcommand{\sO}{{\mfs O}}

\usepackage{color}

\definecolor{blue}{rgb}{0,0,1}
\definecolor{green}{rgb}{0,1,0}
\definecolor{red}{rgb}{1,0,0}
\definecolor{vio}{rgb}{1,0,1}
\definecolor{ama}{rgb}{1,1,0}
\begin{document}

\title{Black hole spectroscopy \\
from Loop Quantum Gravity models}
\author{Aurelien Barrau}
\email{Aurelien.Barrau@cern.ch}
\affiliation{
Laboratoire de Physique Subatomique et de Cosmologie, Universit\'e Grenoble-Alpes, CNRS-IN2P3\\
53,avenue des Martyrs, 38026 Grenoble cedex, France\\
}%
\author{Xiangyu Cao} 
\email{xiangyu.cao08@gmail.com}
\affiliation{
Laboratoire de Physique Th\'eorique et Mod\`eles Statistiques, Universit\'e Paris-Sud, CNRS(UMR 8626), \\91405 Orsay, France\\
}

\author{Karim Noui}
\email{karim.noui@lmpt.univ-tours.fr}
\affiliation{Laboratoire de Math\'ematiques et Physique Th\'eorique, CNRS (UMR 7350), F\'ed\'eration Denis Poisson,Universit\'e Fran\c{c}ois Rabelais, Parc de Grandmont, 37200 Tours, France}
\affiliation{Laboratoire APC -- Astroparticule et Cosmologie, Universit\'e Denis Diderot Paris 7, 75013 Paris, France}

\author{Alejandro Perez}
\email{perez@cpt.univ-mrs.fr}
\affiliation{Centre de Physique ThŽorique, CNRS (UMR 7332)}
\affiliation{Aix Marseille Universit\'e and Universit\'e de Toulon, 13288 Marseille, France}

%
\begin{abstract}
Using Monte Carlo simulations, we compute the integrated emission spectra of black holes  in the framework of Loop Quantum Gravity (LQG).
The black hole emission rates  are  governed by the entropy whose value, in recent holographic loop quantum gravity models, was shown to agree at leading order with the Bekenstein-Hawking entropy. Quantum corrections depend on the Barbero-Immirzi parameter $\gamma$. 
Starting with black holes of initial horizon area $A \sim 10^2$ in Planck units, we
present the spectra for different values of $\gamma$. Each spectrum clearly decomposes
 in two distinct parts: a continuous background which corresponds to the semi-classical stages of the evaporation and a series of discrete peaks which constitutes a signature
of the deep quantum structure of the black hole. We show that $\gamma$ has an effect on both parts that we analyze in details. Finally, we estimate the number of black holes and the instrumental resolution required to experimentally distinguish between the considered models.

  \end{abstract}
%

\maketitle
 
\section{Introduction}

Loop quantum gravity (LQG) \cite{lqg1, lqg2, lqg3} proposes a description of the fundamental degrees of freedom responsible for the
black hole (BH) entropy (see \cite{G.:2015sda} and references therein). According to the most recent results \cite{Engle:2009vc, Engle:2010kt, Engle:2011vf}, these fundamental excitations live on the horizon and are 
elements of the Hilbert space of a Chern-Simons theory: the gauge group is $SU(2)$, 
the canonical surface is  a punctured two-sphere, and the level (coupling constant)
$k$ is proportional to the horizon area $A$ in Planck units.  Punctures are quanta of area associated with horizon-piercing edges of the spin-network-states that define the 
quantum states of the bulk exterior geometry. 

As the quantization of a Chern-Simons theory with a compact gauge group is now well-understood (see \cite{noui1,noui2,noui3} and references therein), 
the kinematical characteristics of a quantum black hole within the framework of LQG are very well-defined. 
This  enables the identification of the microstates for a black hole in equilibrium. The characterization of their number
for a given macroscopic horizon area $A$  allows for the computation of BH entropy and leads to compatibility with the celebrated Bekenstein-Hawking area law
\be\label{one}
S_{BH}= \frac{A}{4\ell_p^2}.
\ee 


Due to the dependence of the area spectrum on the Immirzi parameter $\gamma$ compatibility with Bekenstein-Hawking BH entropy required  fixing the Immirzi parameter 
to a special numerical value in early BH entropy calculations\cite{Rovelli:1996dv, Ashtekar:1997yu}.  More precisely in these models the BH entropy was shown to be
given by
\be
S=\frac{\gamma_0}{\gamma}  \frac{A}{4\ell_p^2},
\ee
where $\gamma_0$ takes an order one numerical value depending on the models \cite{Domagala:2004jt, Ghosh:2004rq}. 
Compatibility with (\ref{one}) was interpreted as a constraint on the value of $\gamma$ imposed by the existence of the correct semiclassical regime. However, 
 the Immirzi parameter is the coupling constant with a topological term in the action of gravity \cite{Date:2008rb, Rezende:2009sv} with no impact in the classical equations of motion.
 Hence, the strong dependence of the black holes entropy computation on $\gamma$ remained a controversial aspect.  For further discussion of this and an exploration of the influence of $\gamma$  in the phase space structure of gravity see \cite{gamma1,gamma2,gamma3}.  
  
A promising perspective on the issue of the $\gamma$ dependence of BH entropy in LQG  was recently put forward thanks to the 
availability of the canonical ensemble formulation of the entropy calculation making use of
 the quasi-local description of black holes \cite{Frodden:2011eb}. 
It was shown in \cite{Ghosh:2011fc} that the semi-classical thermodynamical behavior of BHs can be recovered for all values of 
$\gamma$  if one admits the existence of a non trivial ($\gamma$-dependent) chemical potential
conjugate to the number of horizon punctures.  The problem of the observability of the puncture number is resolved by taking into account  
contributions to the area degeneracy coming from the matter sector.
 As suggested by the semiclassical description of vacuum fluctuations of non geometric degrees of freedom close to the horizon one expects such 
 contributions to degeneracy to grow exponentially with the BH area\cite{Solodukhin:2011gn}. In \cite{Ghosh:2013iwa} one postulates such a phenomenological contribution and shows that compatibility with the existence of large 
 semiclassical black holes implies that the matter sector saturates the holographic bound for a vanishing chemical potential (the number of punctures drops out, in this way, of the list of macroscopic observable quantities).  In such a scenario, the entropy of large semiclassical black holes coincides with (\ref{one}) to leading order, while the dependence on the Immirzi parameter is shifted to sub-leading quantum corrections.
    
A possible fundamental explanation the exponential degeneracy of the assumption made above was  proposed in \cite{Frodden:2012dq} (and developed further in \cite{analytic}).
In these works, the area degeneracy (viewed as an analytic function of $\gamma$) is analytically continued  
from real $\gamma$ to complex $\gamma$  and then evaluated at the special complex values $\gamma=\pm i$
to find that it grows asymptotically as $\exp(A/(4\ell_p^2))$ for large areas. 
The result is striking in that these values of the Immirzi parameter are special in the connection formulations of gravity:
they lead to the simplest covariant parametrization of the phase space of general relativity in terms of the so-called Ashtekar variables \cite{Ashtekar:1986yd}. 
These results suggest that the quantum theory, when defined in terms of self dual variables,  might automatically account for a holographic degeneracy of the area spectrum of the BH horizon.


Given the present landscape of different models, and leaving aside theoretical reasons for preferring one over the others, it would be interesting to test
their differences from an (idealized) observational viewpoint. The value of black hole entropy provides little information allowing for such a distinction
as the models (either by fixing $\gamma=\gamma_0$ or by arguing for an additional exponential degeneracy coming from either semiclassical arguments or 
from analytic continuation to self dual variables) all coincide in reproducing (\ref{one}) to leading order.

The present  work shows that a possible way of distinguishing the models is to analyze out-of-equilibrium processes. More precisely, we describe
the black hole radiation process using statistical mechanical techniques together with some assumptions about the dynamical behavior of BHs close to equilibrium.
In doing this we will basically follow the ideas of \cite{barrau} leading to the computation of the integrated  
emission spectrum for a black hole (under some assumptions which will be made precise later). The simulated spectrography will allow for the study of deviations predicted by the old and new (holographic) models of LQG from the semiclassical Hawking process expectation. This is of special interest for the holographic models of \cite{Ghosh:2013iwa, Frodden:2012dq}
for which the effect of $\gamma$-dependent sub-leading quantum corrections  could be detectable.

The article is organized as follows. In section \ref{II} we recall  basic facts about black hole evaporation. In section \ref{III} we give a brief description 
of the different models for black holes that we study in this paper. In section \ref{IV} we present the numerical methods used for the simulations. 
Results are presented and discussed in section \ref{V}. We finish with a discussion in section \ref{VI}.
From now on we consider units such that $G=c=\hbar=1$. In particular, areas, lengths and energies are given in Planck units.

\section{Evaporation and transition rates}\label{II}
According to the celebrated no-hair-theorem, stationary (axisymmetric) electro-vacuum black holes (those of astrophysical relevance) are described by members of the 
Kerr-Newman family and are labelled by their ADM mass $M$, angular momentum $J$ and electric charge $Q$. The dynamics of the system under small disturbances is described by the first law of BH mechanics
\begin{eqnarray}\label{1st}
\delta M=\frac{\kappa}{8\pi} \delta A +\Omega \delta J+\Phi  \delta Q,
\end{eqnarray}
where $A$ is the horizon area,  $\kappa$ its surface gravity, $\Phi$ is the horizon electric potential and $Q$ is the black hole electric charge. 

Hawking's semiclassical considerations \cite{Hawking:1974sw} imply that Kerr-Newman black holes radiate particles according to the Planck's radiation law at temperature $T=\kappa/(2\pi)$. More precisely, the mean number of particles  $\sN(\omega,m,q)$ 
with momentum  $k^a$ radiated at infinity and carrying energy $\omega$, angular momentum quantum number $m$ and charge $q$  is given by
\ba\label{hawkspec}
\langle \sN({\omega,m,q}) \rangle
&=&
 \frac{\Gamma(\omega, m, q)}{\exp\left(\frac{2\pi}{\kappa}(\omega -\Omega m -q \Phi)\right)\pm 1},
\ea
where $\omega=-k^a\xi_a$ and $m=k^a\psi_a$ are the frequency and angular momentum of the mode at infinity, and $\xi^a$ and $\psi^a$ are the Kerr-Newman stationarity and axi-symmetric Killing vectors respectively normalized at infinity. 
The function $\Gamma$ in (\ref{hawkspec}) is the so-called greybody factor, 
and the $\pm$ sign in the denominator depends on whether one is considering radiation of  bosons ($+$) or fermions ($-$). 

Assuming that when a wave packet carrying energy $\omega$, angular momentum $m$ and charge $q$ is emitted the black hole parameters are readjusted 
according to $\delta M=\omega$, $\delta J=m$ and $\delta Q=q$, we can write,  from the first law (\ref{1st}), the mean number of emitted particles as follows
\ba\label{lili}
\langle \sN  \rangle
&=&
  \frac{\Gamma(\omega, m, q)}{e^{\frac{\delta A}{4}}\pm1},
\label{three}\ea
which would  make the spectrum Planckian in terms of the ``area quanta
$\delta A$" if it was not for the distortion introduced by $\Gamma(\omega, m, q)$. This distortion has 
nothing to do with the thermal properties of the horizon and appears only when describing the radiation from 
the point of view of observers at infinity: $\Gamma(\omega, m, q)$ encodes the backscattering effects on modes 
propagating from the vicinity of the horizon out to infinity in the background Kerr-Newman space-time. 

\subsection{The quasi-local point of view}

From the point of view of stationary observers, denoted $\sO$ in what follows,  at a proper distance $\ell\ll M$ from the horizon, one obtains
\ba\label{eruct}
\langle \sN  \rangle
&=&
  \frac{\Gamma_0}{e^{\frac{\delta A}{4}}\pm1},
\ea
where $\Gamma_0$ is a constant, and thus  
the spectrum becomes strictly Planckian in area quanta.
 As shown in \cite{Frodden:2011eb} the (global) first law (\ref{1st}) transforms into the following simplest local first law for $\sO$:
\begin{eqnarray}\label{1stlo}
\delta E =\frac{\overline \kappa}{8\pi}\delta A,
\label{firstlaw}
\end{eqnarray}
where $\delta E$ is  the local energy changes as measured by $\sO$. The variable $\overline \kappa$ denotes the local surface gravity corresponding to the local acceleration of the stationary observers defined above, {\it i.e.} $\overline \kappa=\|{ a_{\va \sO }}\|$. The (Rindler like) near horizon geometry implies that $\overline\kappa=1/\ell$.

Notice that $\overline \kappa$ is  universal in the sense that it is independent of the macroscopic parameters   
defining the black hole \cite{Ghosh:2011fc}.  An important consequence of this is that one can integrate the  local first law 
to obtain a quasi-local energy notion $E$  associated with the (black hole) system from the perspective of the observers $\sO$, namely 
\begin{eqnarray}\label{quasi local energy}
E=\frac{A}{8\pi \ell}.
\end{eqnarray}
 Applying the first law (\ref{firstlaw})  to the energy variation corresponds to the emission of a particle  with four momentum $k^a$ and local frequency $\overline \omega\equiv k_au^a$ 
as measured by $\sO$ (with four velocity  $u^a$) 
implies that
 \begin{eqnarray}
\frac{ \delta A}{4}= \frac{2\pi \overline\omega} {\overline\kappa},
 \end{eqnarray}
which turns (\ref{lili}) into a  Planckian spectrum at the Unruh temperature 
\be
T= \frac{\overline \kappa}{2\pi}=\frac{1}{2\pi \ell},
\ee 
as measured by $\sO$.
The frequency independence of $\Gamma_0$ in (\ref{eruct}) in the quasi-local treatment follows from the
scale invariance of the near horizon approximation\footnote{In the spherically symmetric case, the Klein-Gordon equation simplifies to a two dimensional wave equation in the transversal coordinates
$t$ and $r_*$ (where $r_*$ is the tortoise coordinate) with an effective potential that vanishes at the horizon \cite{Wald}. This can also be shown in the rotating black hole case.}.

\subsection{Entropy and transition rates}
The integration of the  first law, together with the area theorem, leads to the thermodynamical entropy 
\be\label{entro0}
S = \frac{A}{4}+S_0,
\ee
where $S_0$ is an integration constant that cannot be fixed by a thermodynamical reasoning. 
In fact, as in any thermodynamical system, entropy cannot be determined only by the use of the first law. 
It can either be measured in an experimental setup (this was the initial way in which the concept was introduced) or calculated  by using statistical mechanical methods once a model for the fundamental building 
blocks of the system is available. For instance, by computing the  microcanonical entropy  $S(A)=\log(\sN(A))$ where $\sN(A)$ is the number of black hole microstates compatible
 with the macroscopic horizon area $A$. 

When one takes into account the evaporation phenomenon, the black hole can no longer be considered as a system at thermodynamical equilibrium. However,
for dynamical  systems which remain close to equilibrium at any time during their evolution, the statistical entropy  is still a well-defined quantity:
we say that the system is at a local thermodynamical equilibrium. This would be the case of large black holes for which Hawking temperature is cold and the 
horizon area $A$ evolves slowly. In such close to equilibrium scenarios, the micro states of the horizon compatible with the instantaneous area $A$ can be assumed to be equally likely.
Thus the probability ${\cal P}(A \to A - \delta A)$ for the black hole of horizon area $A$ to 
evaporate a piece of area $\delta A$ is well approximated by the rate of final to initial number of microstate states
\be\label{proba}
{\cal P}(A\to A-\delta A) \propto \frac{\sN(A-\delta A)}{\sN(A)} =  \exp{[-\delta S(A)]}.
\ee
The proportionality factor is fixed by the requirement that the probability density is normalized.
Taking into account backscattering  and normalization issues, one shows that the probability for a black hole to emit a particle whose energy corresponds to the decrease of area $\delta A$ is 
given by
\be\label{area}
{\cal P}(A\to A-\delta A)=\Gamma(A,\delta A)  \exp{[-\delta S(A)]}.
\ee

We will assume the BH to be of Schwarzschild type.
{ Consequently, in the optical limit $M\omega\gg 1$ the greybody factors $\Gamma\approx M^2\omega^2$ independently of the particle's spin. In the other limit  $M\omega\ll 1$ one has $\Gamma \approx A=16\pi M^2\omega^2$ for scalar particles, $\Gamma \approx 2\pi M^2\omega^2$ for spin $1/2$ particles, and  
 $\Gamma \approx \frac{12}{9} A M^2 \omega^4$ for photons \cite{Page:1976df}.  
 { Numerical limitations constrain us to study BHs of a maximum area of about $10^2$ Planck areas.
Such BHs have a radius of about $2$ in Planck units which means that the local curvature at their horizon is really Planckian. 
If we boldly stretch the assumption of 
the validity of the semiclassical description of space-time outside the black hole, simple dimensional analysis implies that most of the radiated particles fall into the
$M\omega< 1$ regime. 
}
   

In order to isolate the quantum effects associated with variations of the Barbero-Immirzi parameter it will be convenient (in some cases) to 
assume $\Gamma(A,\delta A)=\Gamma_0$ with
a constant fixed from normalization issues only. Such a choice of $\Gamma$ corresponds to neglecting 
backscattering in the description of the emission process and could be regarded as describing the physics of the local observers $\sO$ placed close to the horizon.
We will often use this terminology in what follows. However, one has to keep in mind that while the quasi local perspective is very useful in thermodynamical investigations, the notion of
spectrography does not really make sense for $\sO$. The reason for this is that quasi local observers perceive their environment at thermal equilibrium at the Unruh temperature and are, in this sense,
incapable of discerning particles coming from the BH from others in the thermal bath. 
}

\section{Three models of Black Holes}\label{III}
According to (\ref{area}), the evaporation process is completely governed by the area spectrum of the black hole and its entropy.
Here we briefly review the main models in the literature whose spectrography will be studied in section {\ref{V}}.

\subsection{Models where $\gamma$ must be fixed to $\gamma_0$}
 The micro canonical computations taking into account 
 only the quantum geometry excitations with no extra degeneracy associated to non-geometric degrees of freedom ({\it e.g.} matter fields vacuum fluctuations)
 lead to the following general expression for the entropy \cite{Agullo:2009eq}
\be\label{lqg}
S=\frac{\gamma_0}{\gamma} \frac{A}{4} + {o}(A),
\ee
where $\gamma_0$ is a numerical factor of order one depending on the details of the state counting \cite{G.:2015sda}. The quantum corrections $o(A)$ are usually logarithmic at the leading order.
As explained in the introduction, compatibility with the first law imposes the constraint  $\gamma=\gamma_0$. The model only make sense for that special value of the Immirzi parameter. There is therefore no interesting physics associated with varying $\gamma$.  
 

\subsection{The holographic models}

The previous results correspond to the simplified situation where purely gravitational degrees of freedom are taken into account  in the counting. This has been traditionally the (rather artificial) play-ground for testing the theory without worrying about the difficult issue of matter coupling. If one uses the qualitative behaviour of matter degeneracy suggested by standard QFT with a cut-off at 
the vicinity of the horizon ({\it i.e.} exponential growth of vacuum entanglement in terms of the BH area), then the entropy becomes 
\be\label{sqrt corrections}
S=\frac{A}{4}+ \sqrt{\frac{\pi A}{6\gamma}} + o(\sqrt{A}).
\ee
The Immirzi parameter $\gamma$ affects only quantum corrections. This result is in agreement with semiclassical expectations  \cite{Ghosh:2013iwa}. 
Notice that the entropy has been computed using the quasi-local point of view, the  second order corrections to the entropy are explored in \cite{Azin}. 
 
The presence of the $\sqrt{A}$ correction can be understood directly in the microcanonical framework as follows. In the holographic model, the number of black hole
microstates is given by
\begin{eqnarray}\label{micro holographic}
{\cal N}(A) = \exp(\frac{A}{4}) \, p(A),
\end{eqnarray}
where $p(A)$ denotes the number of ways to write $A$ as the finite sum
\begin{eqnarray}
A=8\pi \gamma \sum_{j} n_j \sqrt{j(j+1)},
\end{eqnarray}
where $j$ are distinct half-integers and $n_j$ are integers. At the semi-classical limit (large $A$), the large spins $j$ dominate and then $p(A)$ is nothing but the number of
partitions of the integer $N_A=A/(4\pi \gamma)$.  Its asymptotic has been known for almost a century, namely
\begin{eqnarray*}
\log p(A) = \pi \sqrt{\frac{2N_A}{3}} + {\cal O}(\log N_A) = \sqrt{\frac{\pi A}{6\gamma}} + {\cal O}(\log A),
\end{eqnarray*}
which leads immediately to the quantum corrections of the entropy  (\ref{sqrt corrections}). In the present context one does not need to fix the value of the Immirzi parameter
and there is interesting physics associated to its variations. We will study the effects of varying $\gamma$ on the BH spectrography in section \ref{V}.

 As explained in the introduction, a fundamental understanding of the holographic hypothesis in the models leading to (\ref{sqrt corrections}) 
 might come from the recent results \cite{Frodden:2012nu, Frodden:2012dq, analytic} that indicate a relationship between holography and 
 LQG  for complex Ashtekar variables ($\gamma=\pm i$). Nonetheless, if in the holographic treatment the degeneracy originates from the matter fields degrees of freedom, it clearly has a (quantum) geometrical origin in the context of complex variables
  treatment of black holes. One can however  argue that matter and geometry, which are already intimately linked at the classical level,  could give exactly the same contribution to the black hole degeneracy (only does the point of view change).

 \section{Numerical method}\label{IV}
To study the evaporation process and to analyze the effects of the different parameters,  we have developed a Monte Carlo simulation procedure. 
The simulation has been adapted from the one used  in \cite{barrau}. 
The simulation starts at a given mass, whose value is high enough to be well above the deep quantum regime but small enough so that all states below this mass can be explicitly calculated. 
In this paper, we start with black holes of area $A_{\text{init}} = 4\pi\gamma_{\min} \times 70$  with $\gamma_{\min} = 0.2$ which gives $A_{in}=1.76 \times 10^2 $ in Planck units, whatever the value of $\gamma$.  This corresponds to an initial BH mass of the order of $2$ Planck masses, i.e.
\be\label{masstiny}
M_{BH}\approx 2.
\ee
Starting with the same area enables us to make easy comparisons. $10^7$ black hole evaporations are simulated to produce each of the spectra shown in Figure \ref{spectrum}. \\

In each case, a standard-model particle is randomly selected at each step, with a probability weighted by its number of internal degrees of freedom and the corresponding (spin-dependent) greybody factor.  Only photons, neutrinos, and charged leptons are finally kept for the analysis. Quarks and gluons undergo fragmentation and generate hadrons with wide energy distributions. Were new particles beyond the standard models to be emitted, this would not change drastically the results presented here. Only the percentage of mass emitted in the photon-neutrinos-charged-leptons chanel  would change and therefore the number of black holes required to reach a given confidence level. But the shape of the spectra and the differences between the models would qualitatively remain the same.

Once the particle type is selected, its energy is given by the energy loss of the black hole, {\it i.e.} $E=\delta M$. From the (usual) point of view of an observer at infinity,
 the mass $M$ and the horizon area of the black hole are related by $A= 16 \pi M^2$, and then the energy of the emitted particle measured by such an observer is given by
 \begin{eqnarray}\label{infinite vp}
 E=  \delta M =  \frac{1}{4 \sqrt{\pi}} \delta \left(\sqrt{A}\right) \approx  \frac{1}{8\sqrt{\pi}} \frac{\delta A}{\sqrt{A}}, \,
 \end{eqnarray}
where $\delta A$ represents the variation of area due to the emission of the particle. The last identity is true only for small variations $\delta A$
compared to the original area $A$.
The relation between the energy and $\delta A$ is simpler from the quasi-local point of view. In that scheme, we saw that the energy of the black hole is directly proportional to the horizon area according to (\ref{quasi local energy}).
If we denote by $E_\ell$  the quasi-local energy of the emitted particle,
with $\ell$  the distance of the quasi-local observer to the horizon, we obtain immediately
\begin{eqnarray}\label{local vp}
E_{\ell} =  \frac{\delta A}{8\pi \ell} .
\end{eqnarray} 
From a physical point of view, it is more satisfying to plot the spectrum as a function of $E_\infty$; but it is much easier to analyze the results
when the spectrum is given as a function of $E_\ell$. For these reasons, we will give the two representations of the black hole spectrum. \\

The probability of transition between two black hole states of different masses (or areas) is taken to be given by the exponential of their entropy difference weighted by the greybody factor. 
Most probably, those greybody factors should receive quantum gravity corrections, especially for the late stages of the evaporation. These corrections may become very large and  considering only a semi-classical
greybody factor may then lead to over-simplified physical predictions.  
The black hole can undergo a transition to any state having a lower mass.
 In the quantum gravity model we are considering, only discrete values of the area  are possible, and the probability is  driven by the entropy associated with the number of states.
To make comparisons with the semi-classical black hole evaporation, we have also computed numerically the Hawking spectrum which is governed by the classical Bekenstein-Hawking area law.

It should be noticed that, in  quantum gravity models, the transition to the last state, that is $M=0$, naturally happens. In the semi-classical Hawking case, however, the standard formula has to be slightly modified. The Hawking law naively applied would lead the black hole to emit, at the last stage, more energy than it has. We have therefore used a truncation and considered that the energy of the emitted particle is $M$ each time the Hawking spectrum would have led to $E>M$. This only affects the very last emission.\\

The Monte Carlo is by construction a random process. It means that simulating the same process in the same conditions would lead to another spectrum, as in real life, due to the fundamentally quantum nature  of the underlying physics. The spectra presented here are ``mean" spectra corresponding to many evaporating black holes ($10^7$). As the number of emitted particles in the deep quantum gravity regime is quite small (generically less than 10) for a single black hole, obviously only a statistical analysis can lead to a  significant conclusion. To remain realistic, we have added an `experimental' uncertainty on the reconstructed energy of the detected particles. If the resolution was infinite, a single detection would immediately allow to distinguish between the Hawking case (that is a continuum of states) and the quantum gravity models (the distinction between quantum gravity proposals having the same area eigenvalues but different number of micro-states is anyway more subtle). This is however not realistic and we have taken into account a kind of reasonable detector uncertainty. The results presented depend on the number of black holes and on the experimental resolution (see figure \ref{compare}).

\section{Results: black holes spectra}\label{V}
\begin{figure*}
\centerline{\hspace{0.5cm} \(
\begin{array}{c}
\includegraphics[width=7.5cm]{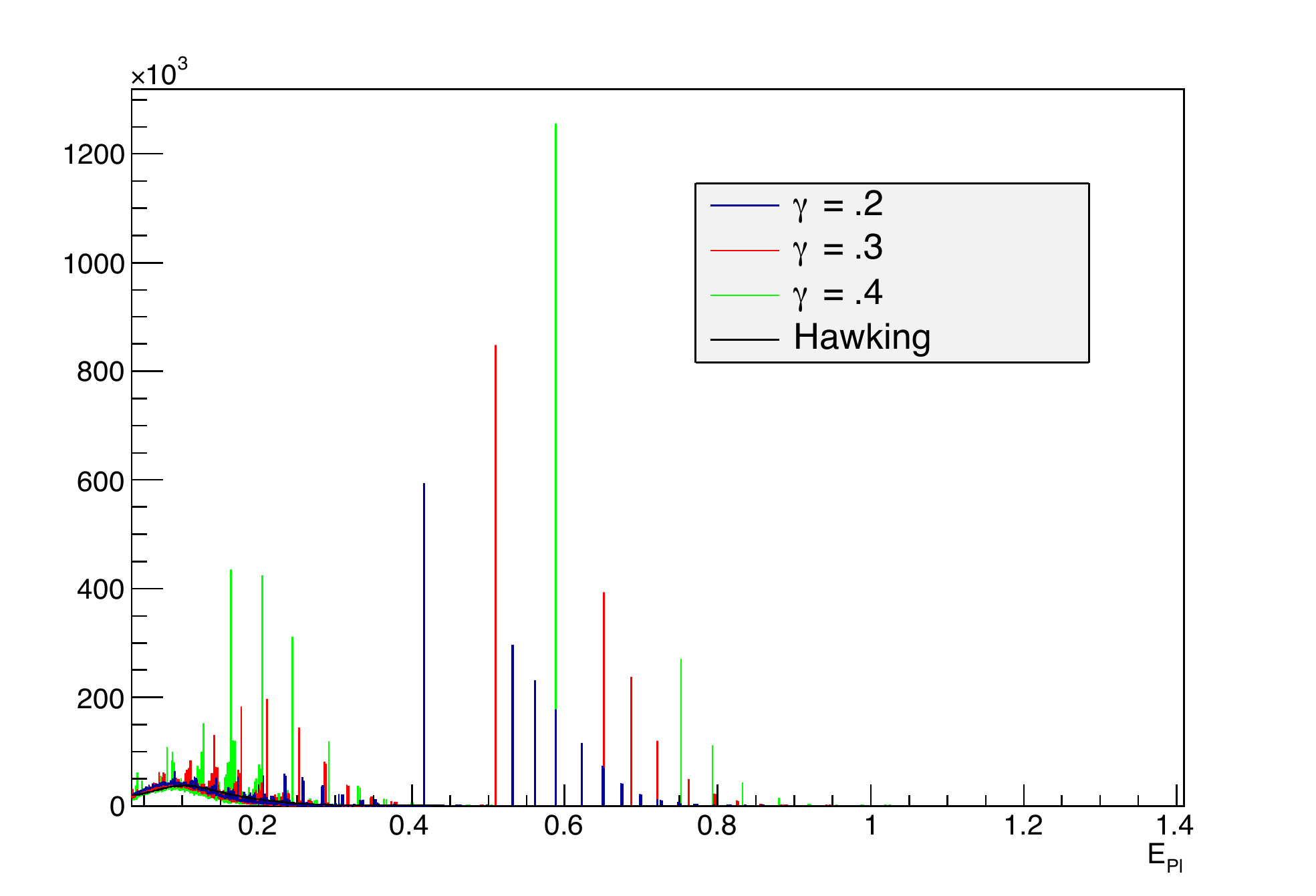}
\includegraphics[width=8cm]{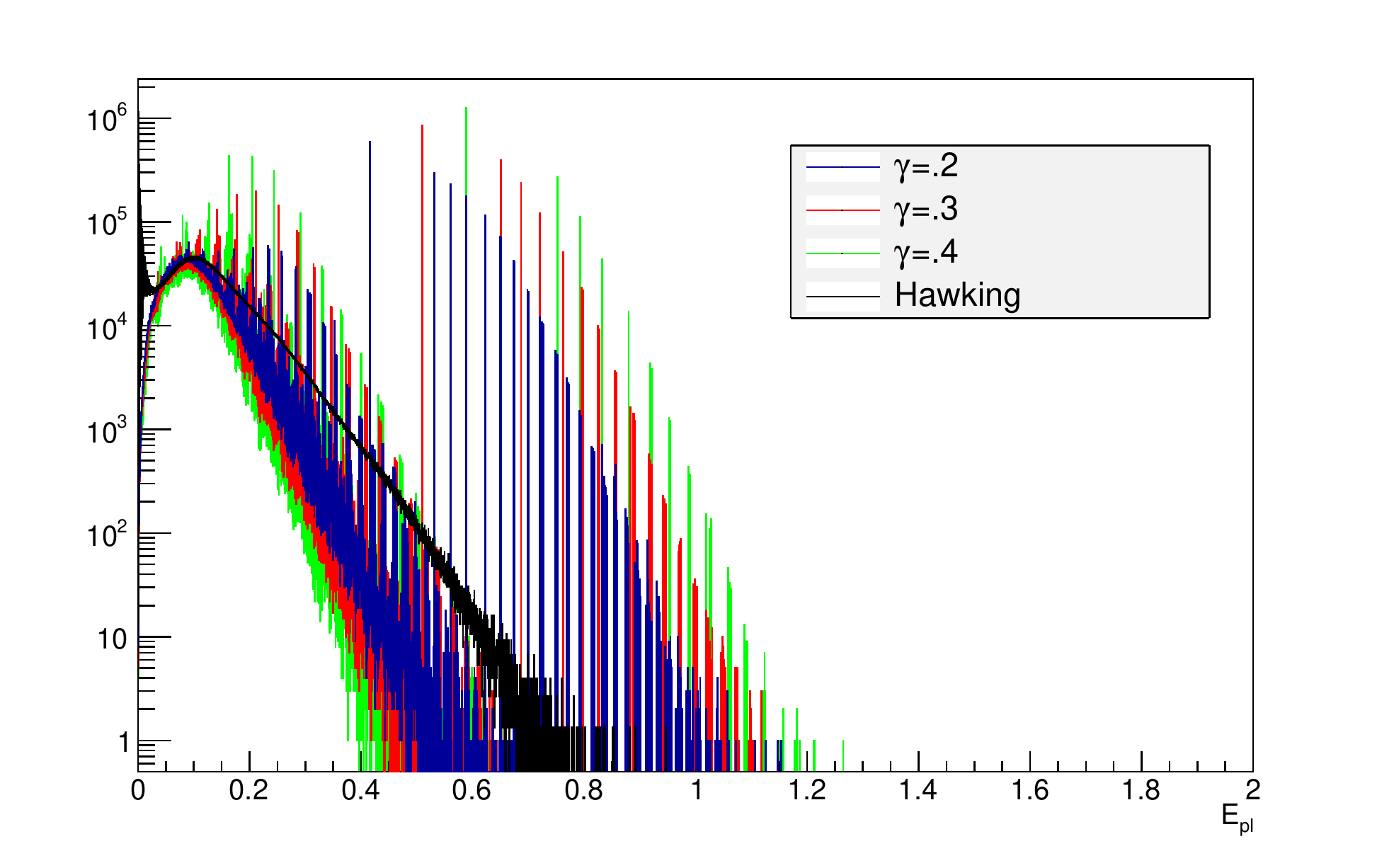}\\
\includegraphics[width=8cm]{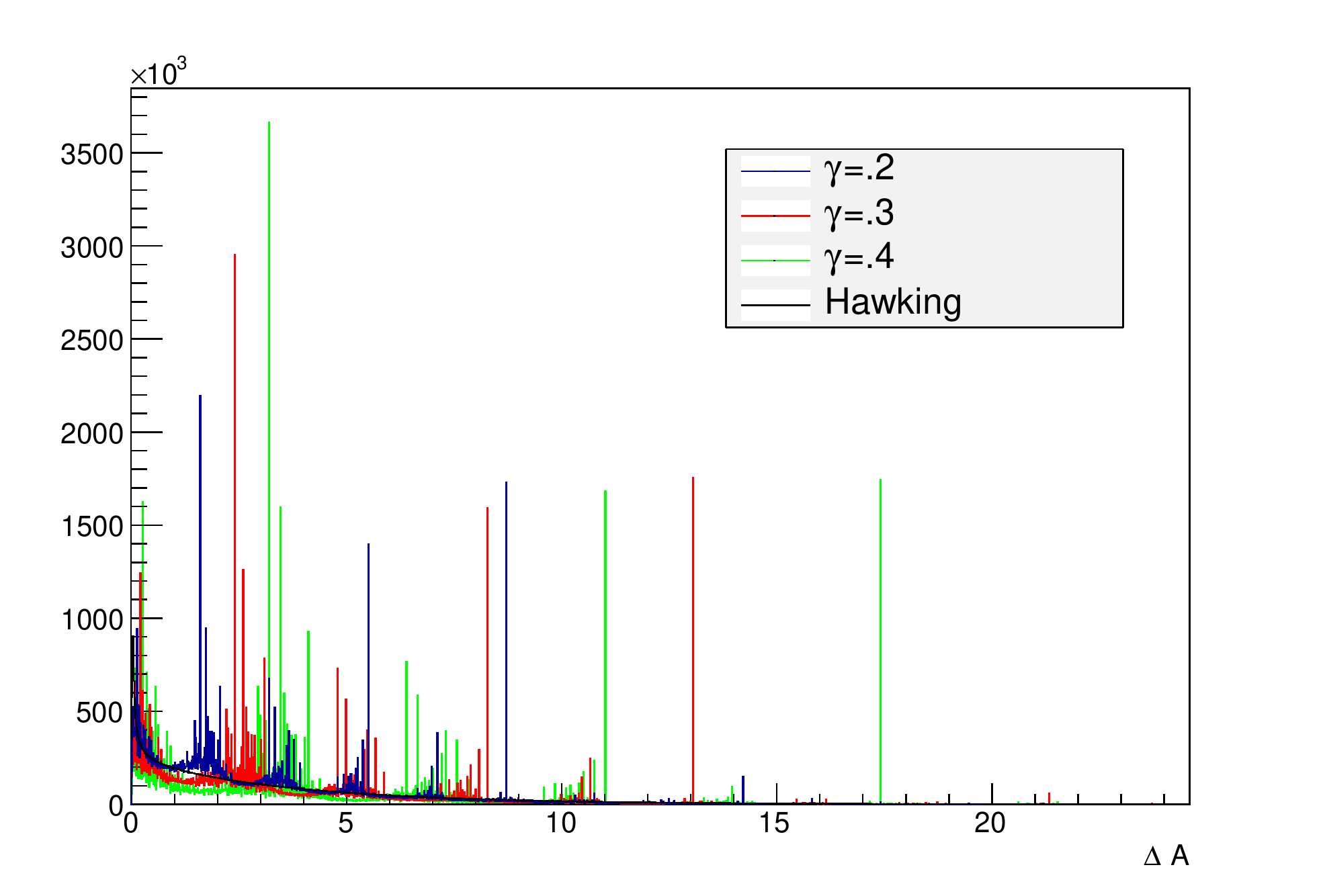}
\includegraphics[width=8cm]{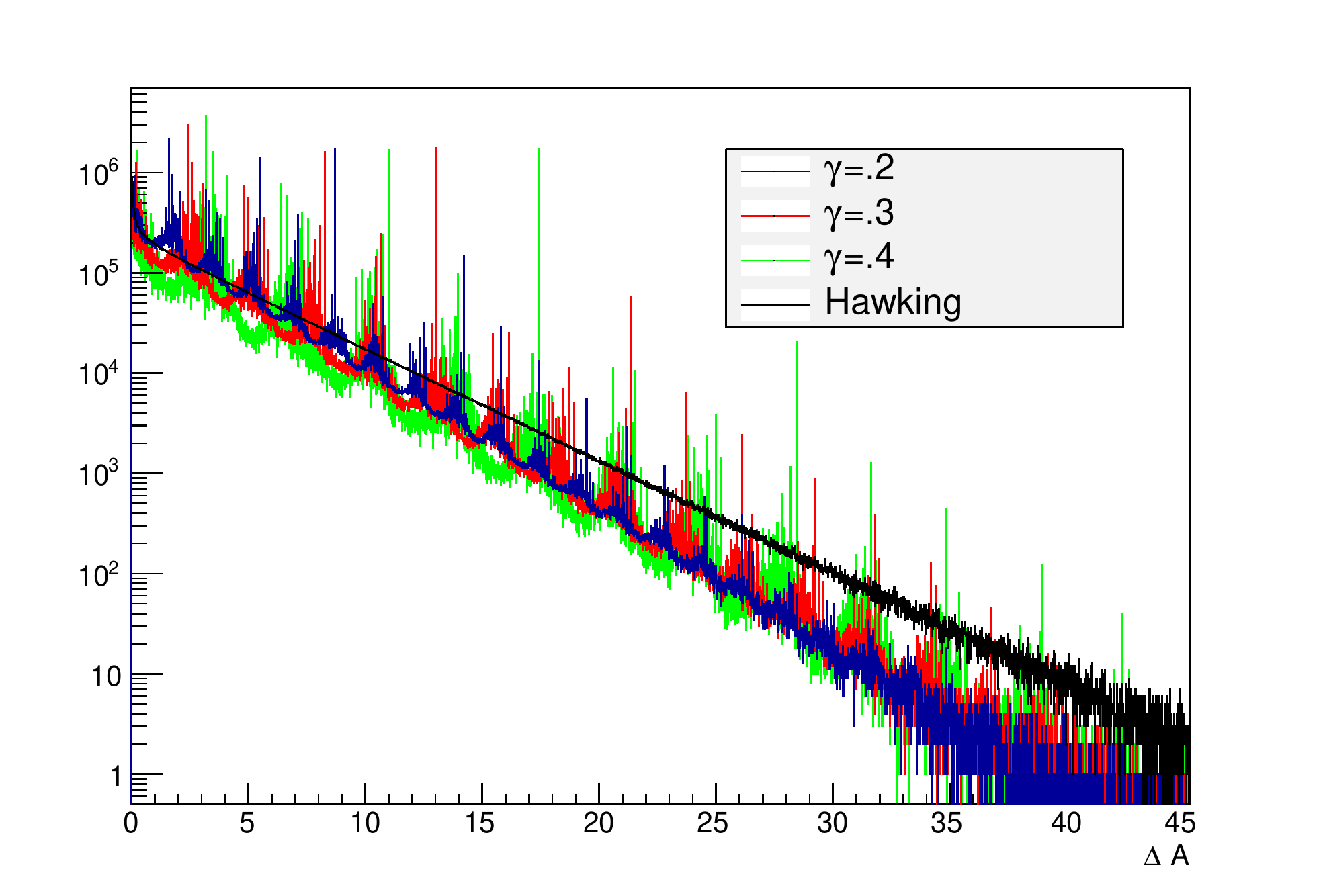}
\end{array}
\)}
\caption{Spectrum of a holographic black hole for different values of $\gamma$.  The spectrum is represented in linear coordinates as a function of $E_\infty$ in the graph at the top left;
it is represented  in logarithmic coordinates as a function of $E_\infty$ in the graph at the top right; 
it is represented in linear coordinates as a function of $\Delta A$ in the graph at the bottom left;
it is represented in linear coordinates as a function of $\Delta A$ in the graph at the bottom right.}
\label{spectrum}
\end{figure*}

We concentrate our analysis on the holographic model where the number of microstates is exactly given by (\ref{micro holographic}). At the semi-classical limit, the resulting entropy
is (\ref{sqrt corrections}). As the greybody factor is fixed to a constant,  the probability for the black hole to emit a particle associated with the decrease $\delta A$ of the horizon area is totally governed 
by the entropy according to (\ref{proba}). In the quasi-local point of view (\ref{local vp}), such a probability amplitude can be expressed in terms of $\delta A$ and is directly related to the black hole spectrum because
the energy of the particle $E_\ell$ is proportional to $\delta A$.  From the point of view of an observer at infinity, the spectrum stems from the probability ${\mathcal P}(A,E)$
which, as we will argue below, takes the form
\begin{eqnarray}
{\mathcal P}(A,E) \propto  \, E^2 \, \exp[-\delta S(A) ], \;
\end{eqnarray}
with the constant fixed by normalization issues.

Using the Monte-Carlo simulation we have just described in the previous sections, we obtained:
\begin{enumerate}
\item exact spectra with different values of $\gamma$  of order $10^{-1}$ (Fig. \ref{spectrum}).  We considered the following set $\{0.2;0.3;0.4\}$ of values for $\gamma$.  
We have ploted the spectra in terms of $\delta A$ (or equivalently in terms of the local energy $E_\ell$)  and also in terms of $E_\infty$. We have presented the results 
both in linear  and logarithmic representations.
\item an accurate comparison of the spectra with the classical Hawking spectrum.
\item the smeared spectra which take into account an instrumental resolution (Fig.\ref{smeared spectrum}).  
\end{enumerate}
Note that the choice of the values of $\gamma$ are motivated by the value for $\gamma$ obtained from old models ($\gamma\approx 0.2$).
Such a choice allows for an easier comparisons  with previous results.

Let us start with some general observations.
Figure \ref{spectrum} shows a strong dependence of the spectrum on the Barbero-Immirzi parameter even though the leading order term of the black hole entropy (which mainly governs the evaporation process) 
is independent of $\gamma$. This dependency, which is completely quantum gravitational in nature, originates from two aspects: $\gamma$ enters in the discretization of the area spectrum and $\gamma$ shows off in the sub-leading corrections to
the entropy. Before going deeper into the explanation of the role of $\gamma$, let us comment further on the general aspects of the spectrum. \\

The spectrum of emission, as seen from infinity, separates into two rather distinct parts: a continuous background whose amplitude is most important in the infrared and a series of discrete peaks which go from the infrared to the deep ultraviolet regime. This separation is  apparent in the top two panels in figure $\ref{spectrum}$ while it is hidden in the bottom panels because the range of the local emission plots in terms of $\Delta A$ has been chosen to illustrate the continuum region of the spectrum only. Notice that the greybody contribution, the factor $E^2$ in the probability amplitude, is responsible for the peak in the emission spectrum as seen from infinity in the continuous lower energy region. As the greybody factors are trivial in the local framework we see a monotonic linear behavior in the logarithmic plot on the bottom right panel.
Coming back to the bottom panels, we observe that the quantum peaks are mainly present
in the deep ultraviolet regime. While the continuous part of the spectrum corresponds to particles emitted at the
early stage of the evaporation process, the peaks reveal the deep quantum structure of the black hole corresponding to the latest stages of the evaporation when the discrete structure of the area is no longer negligible.  These two regions of the spectrum exhibit structures that are dependent on the Barbero-Immirzi parameter. We will now describe these features in more detail.

\subsection{The continuous background: link to the semi-classical evaporation}

In both the quasi-local and the infinity points of view,  spectra have a continuum background which is exponentially decaying.

\subsubsection{The quasi-local point of view: analysis of $\text{Sp}(\Delta A)$}
This aspect is best observed in the logarithmic scale when the spectrum is 
viewed as a function of $\Delta A$. In the infrared, the spectrum can be modeled, in a first approximation, as follows:
\begin{eqnarray}
\text{Sp}(\Delta A) \simeq C \exp (-\tau \Delta A),
\end{eqnarray}
where $\tau$ is the exponential decay rate and $C$ is independent of $\Delta A$. It is interesting to observe that $\tau$   depends on the Barbero-Immirzi parameter 
(and probably  also on the normalization constant $C$ but this aspect is less interesting). We clearly see that $\tau$ decreases with $\gamma$.

All these observations are in fact easy to interpret.  Indeed, in the infrared, the spectrum can be reasonably approximated by its semi-classical expression
which is given by the probability rate ${\cal P}(A \rightarrow A-\Delta A)$. This approximation immediately implies the following estimate for $\tau$:
\begin{eqnarray}\label{tau}
\tau \simeq \frac{\delta S}{\delta A} = \frac{1}{4} \left( 1 + \sqrt{\frac{2\pi}{3\gamma A}}\right),
\end{eqnarray}
where we used the semi-classical expression for the entropy (\ref{sqrt corrections}) up to the first quantum correction.
Here $A$ can be identified with the initial black hole area, {\it i.e.} $A = 1.76 \times 10^2$, as the continuous spectrum concerns the early stage of the evaporation
process when the black hole area remains close to its initial value. The previous expression for $\tau$ is consistent with the numerical simulations. 

We can go further in the analysis of $\text{Sp}(\Delta A)$. We notice that 
the exponential decay approximation is clearly better in the infrared than 
in the ultraviolet where, even if the statistical fluctuations are more important,
the slope $\tau$ seems to increase with $\Delta A$: the spectrum is steeper for large $\Delta A$ (larger than 30 in Planck units) compared to small $\Delta A$ (between 0 and 10 Planck units). 
The previous expression for $\tau$ (\ref{tau})
allows us to easily interpret this fact. In the ultraviolet, the area $A$ in (\ref{tau}) can indeed no longer be identified with the initial area because we are far from the earliest stages of the evaporation process. In that case, $A$ has to be identified
with the instantaneous area of the black hole which is obviously smaller than the initial one. The consequence is that the slope $\tau$ increases because it scales as $1/\sqrt{A}$. The more we go to the ultraviolet, the larger the slope is, exactly
as  observed in the numerical simulations. 

Let us finish the analysis of $\text{Sp}(\Delta A)$ with some remarks concerning the comparison with the Hawking spectrum. We observe that 
the quantum and the Hawking spectra are closer one to the other in the infrared than in the ultraviolet. The reason is simple and comes from the fact that quantum corrections to the entropy
are no longer negligible in the late stages of the evaporation. From the expression of the entropy, we see that the larger $\gamma$ is,  the smaller  quantum corrections are and then the better the approximation to 
the Hawking spectrum in the ultraviolet is. This theoretical prediction is consistent with the numerical calculations (see Figure \ref{spectrum}). 

In the infrared, the situation is somehow the reverse. In that case, the quantum corrections to the BH entropy are negliguible,
and thus cannot explain the differences between the spectra with different $\gamma$. We observe that the smaller $\gamma$ the better the approximation to the Hawking spectrum. This simply comes from the value of the area gap in the 
area spectrum of the black hole which is linear in $\gamma$. When $\gamma$ is small, the area spectrum of the black hole resembles a continuous spectrum and then we expect the emission spectrum to be very close to the Hawking one,
as shown by  the numerical simulation.

\subsubsection{The  point of view at infinity: analysis of $\text{Sp}(E)$}
\begin{figure}[h!!!!!] \centerline{\hspace{0.5cm} \(
\begin{array}{c}
\includegraphics[width=9cm]{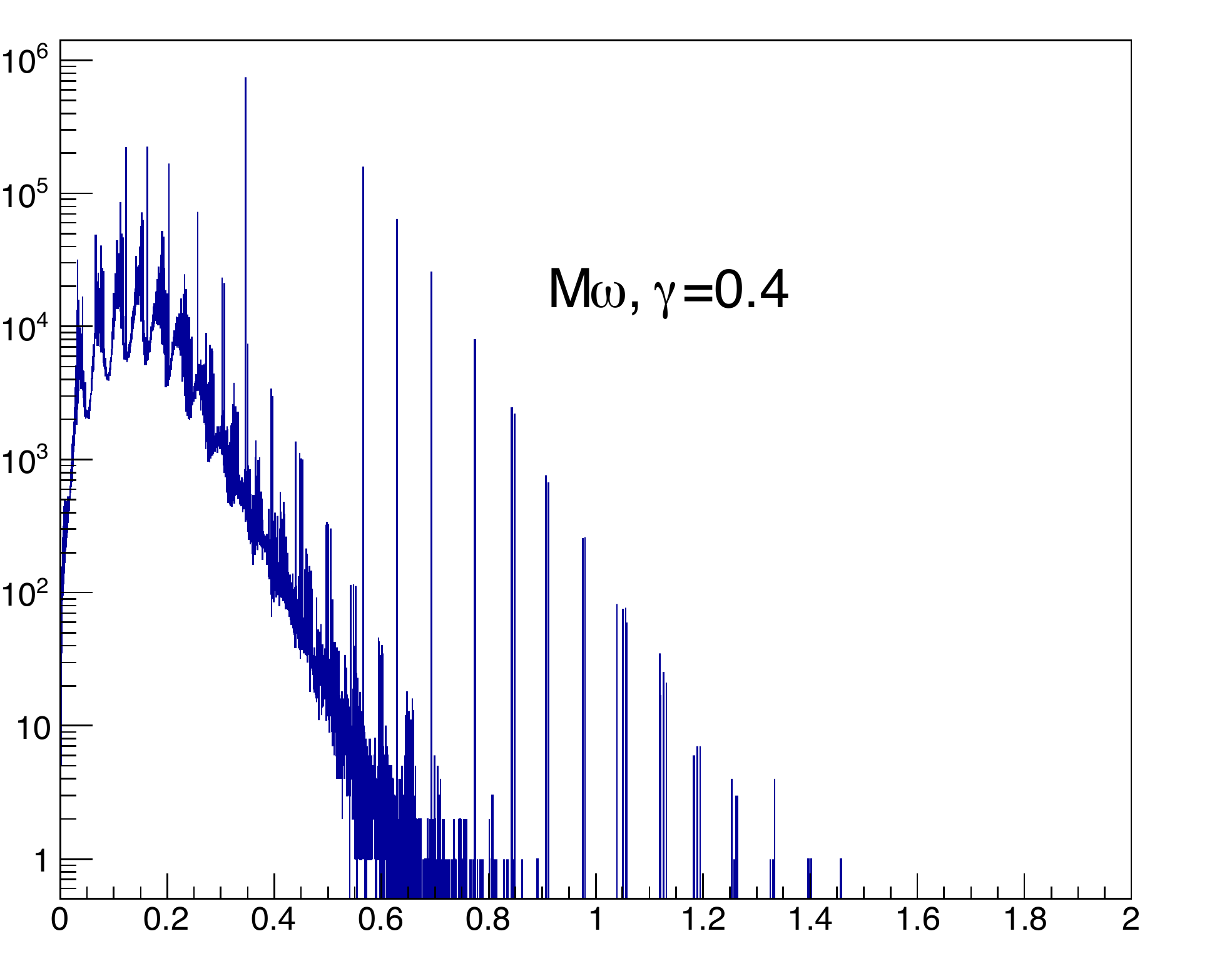}
\end{array}
\)}
\label{footyni}
\caption{Distribution of $M\omega$ for $\gamma=0.4$ illustrating the fact that the continuous part of the spectrum is radiated in the IR regime 
$M\omega<1$.} \label{old1}
\end{figure}

When the spectrum $\text{Sp}(E)$ is viewed as a function of $E$ (which is relevant for the point of view of an observer at infinity), we observe that its continuous background admits a maximum and the smallest energies are suppressed. 
From a theoretical point of view, this is explained because, near the maximum, the continuous background can be approximated by its semi-classical expression which leads to the following estimate\footnote{As the initial mass of the BH $M_{in}\approx 2$ the continuous part of the spectrum falls in the regime $M\omega\ll 1$ (putting all the units in the previous criterion gives $(M(\hbar \omega/c^2))/M_p^2=(ME)/M_p^2<1$). 
In the Figure \ref{footyni} below we show the exact distribution of $M\omega$ for $\gamma=0.4$.
This means that we can use the low energy form of the form factors \cite{Page:1976df} from which we would get  \ba \n(N_{f} a_f E^2+N_{ph} a_{ph} E^4) \exp \left( - \tilde{\tau} E\right)\approx\\  N_{f} a_f E^2 \exp \left( - \tilde{\tau} E\right),
\ea
where $N_f$ and $N_{ph}$ are the number of fermions and photon degrees of freedom and $a_f$ and $a_{ph}$ are parameters of the same order. 
The photon term is subdominant as $N_f> N_{ph}$ in the standard model. } 
\begin{eqnarray}
\text{Sp}(E) \simeq \tilde{C} E^2 \exp \left( - \tilde{\tau} E\right),
\end{eqnarray}
where the slope $\tilde{\tau}$ is now given by
\begin{eqnarray}
\tilde{\tau} \simeq \frac{\delta S}{ E} \simeq 2\sqrt{\pi A} + \sqrt{\frac{8 \pi^2}{3\gamma}}.
\end{eqnarray}
As in the previous analysis, $\tilde{C}$ can be approximated by a constant and $A$ can be reasonably identified with the initial area. From this result, we 
can immediately estimate the energy $E_{max}=2/\tilde{\tau}$ for which the spectrum is maximum together with the  value of this maximum which is given by
\begin{eqnarray}
\text{Sp}(E_{max}) \simeq \tilde{C} \left( \frac{2}{e \tilde{\tau}} \right)^2. 
\end{eqnarray}
Therefore, we can predict that the maximum of the spectrum increases with $\gamma$. This is exactly what we observe in the numerical experiments.
Such an analysis shows that the quantum corrections to the entropy manifest themselves also in the infrared part of the spectrum, and not only in the deep 
ultraviolet regime, what one would have expected. This aspect  seems to us very interesting to underline.

\subsection{The quantum peaks: quantum nature of space-time}
 In Figure \ref{spectrum}, the linear plot ($\text{Sp}(\Delta A)$ or $\text{Sp}(E)$) is the more convenient to analyze and interpret the  properties of the quantum peaks in the 
 black hole spectrum.
These peaks (above all the ones with a higher energy) are expected to appear at the late stages of the black hole evaporation process and therefore should correspond to the transitions between low area states. 
In that respect, they provide us with a concrete signature of the  deep quantum structure of the black hole. 

More explicitly, each peak is produced by a transition from a state $i$ (of quantum area $A_i$) to a state $j < i$ (of quantum area $A_j$). Whether  we consider the local point of view or the point of view at infinity,
the energy of the corresponding emitted particle is given by
\begin{eqnarray}
E_{\ell ij}  & =  & \frac{\delta A_{ij}}{8\pi \ell} \propto \gamma, \\
E_{\infty ij} & = & \frac{1}{4\sqrt{\pi}}(\sqrt{A_i} - \sqrt{A_j}) \propto \sqrt{\gamma}, \,
\end{eqnarray} 
and therefore it scales respectively as $\gamma$ or $\sqrt{\gamma}$. This scaling can be immediately observed in the linear plot $\text{Sp}(\delta A)$ and $\text{Sp}(E)$. For instance, the three highest peaks in $\text{Sp}(\delta)$
correspond to the following approximate energies:
\begin{eqnarray}
\begin{array}{|c|c|c|c|}
\hline
\gamma & 0.2 & 0.3 & 0.4 \\
\hline
\delta A(\gamma) & 1.60  & 2.40  & 3.20  \\
\hline
\delta A /\gamma & 0.80  & 0.80 & 0.80 \\
\hline
\end{array}
\end{eqnarray}
Indeed, $A(\gamma) = 4\pi \gamma \left( 2\sqrt{1\times3} - \sqrt{2 \times 4} \right)$, which is the area difference between any configuration containing (at least) two $j = \frac{1}{2}$ punctures and the configuration obtained by replacing them by one $j = 1$ puncture. All other area--difference peaks can be identified to a puncture change or removal in the same manner. \\

The same exercise for highest energy peaks can be done and the observation of the scaling of the energy as $\sqrt{\gamma}$ in the point of view at infinity can be checked. Nevertheless, since $\delta E$ depends on $\delta A$ \textit{and} $A$, peaks are to be associated with transition between two configurations. For instance, the highest peaks correspond to the transition $(n_j) = (2,0,0\dots) \rightarrow (n_j) = (0,0,\dots)$ with $E_{\infty} = \sqrt{\gamma} 3^{1/4} 2^{-1/2}$, which gives $0.42, 0.51, 0.59$ for $\gamma = 0.2, 0.3, 0.4$ respectively. 

A more delicate effect of $\gamma$ concerns the  amplitude of the peaks. This effect is easier and more precise to study in the quasi-local point of view where the amplitudes
are given by $\text{Sp}{(E_{ \ell ij})}$. In fact, to take into account the normalization issues, it is more judicious to study relative amplitudes of the form $\text{Sp}{(E_{ \ell ij})}/\text{Sp}{(E_{\ell0})}$
where  $E_0$ is the reference energy corresponding to a given transition $i_0$ to $j_0$. One can for instance choose the highest energy peak as the reference. Such relative amplitudes depends on $\gamma$ according to
\be
\frac{\text{Sp}(E_{\ell ij})}{\text{Sp}(E_{\ell 0})} \propto \exp [ - 2\pi \ell (E_{\ell ij} - E_{\ell 0})].
\ee
As the energy scales as $\gamma$ in the quasi-local point of view, the relative amplitude is more suppressed for larger $\gamma$.
This explains the observation that when $\gamma$ increases, lower energy peaks have larger amplitudes. In the infinite viewpoint, although it is not\textit{ a priori } evident, we observe that the presence of greybody factors do not alter the qualitative picture in the range of $\gamma\in [0.2, 0.4]$ considered here. We leave the issue of how these features extrapolate to a larger range of $\gamma$ for further studies, the main message here being that the effect of $\gamma$ on the spectrum (in both its semi-classical and deep quantum regimes) is more than a simple dilation.
  
\subsection{Comparison with older models}
\begin{figure}[h!!!!!] \centerline{\hspace{0.5cm} \(
\begin{array}{c}
\includegraphics[width=9cm]{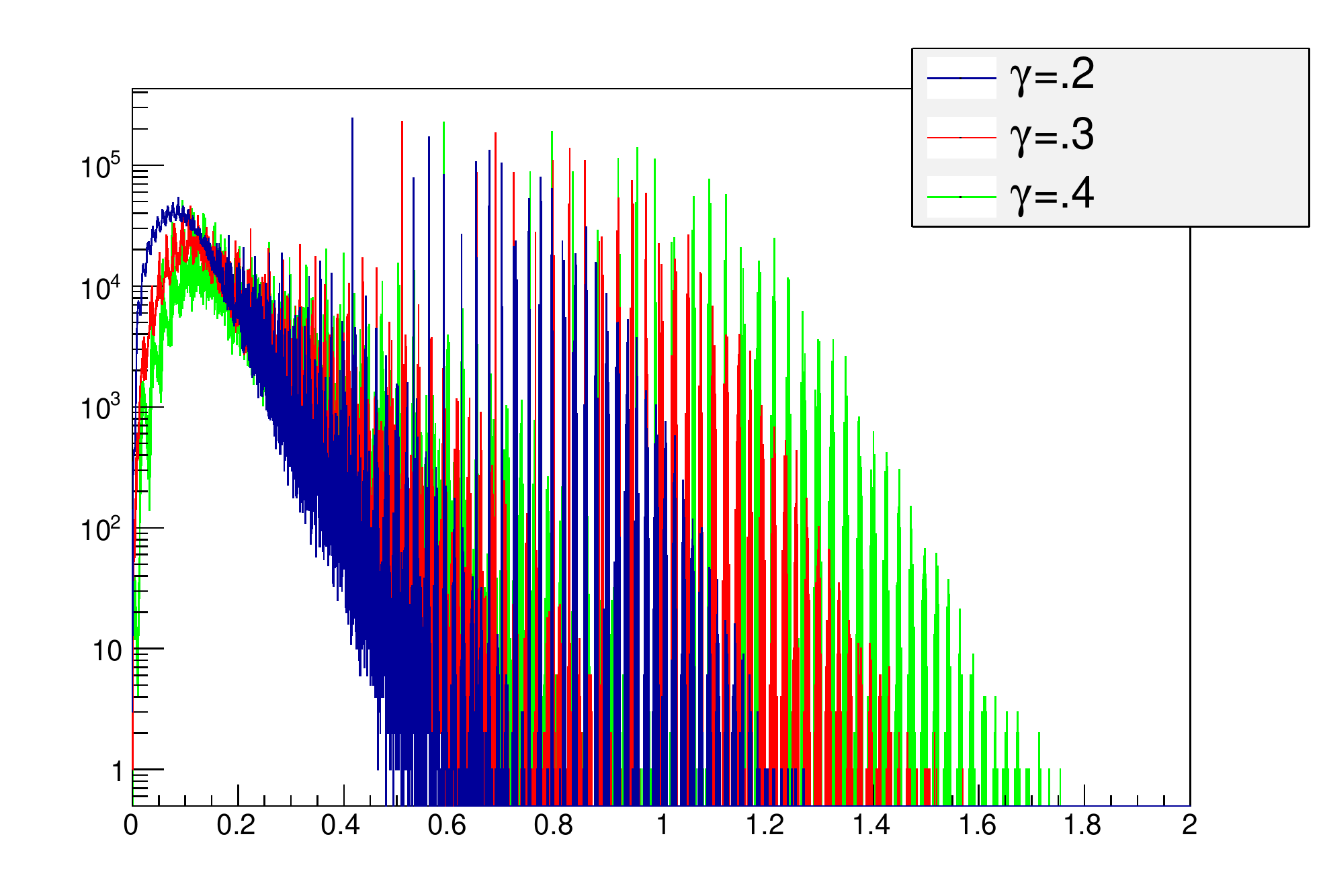}
\end{array}
\)}
\label{oldylog}
\caption{$\gamma$ dependence of the integrated spectrum in the standard LQG model (logarithmic scale).} \label{old2}
\end{figure}

For completeness we here show in figures  \ref{old1} and \ref{old2}  the $\gamma$ dependence of the integrated spectrum of the non-holographic models considered in previous literature \cite{barrau}.
Notice that a comparison of different values of $\gamma$ is only a mathematical exercise as in the framework of such models the Immirzi parameter must be fixed
to the value that is compatible with $S_{BH}=A/4$ which in this case corresponds to $\gamma\approx0.274$ \cite{G.:2015sda}. 
\begin{figure}[h!!!!!] \centerline{\hspace{0.5cm} \(
\begin{array}{c}
\includegraphics[width=9cm]{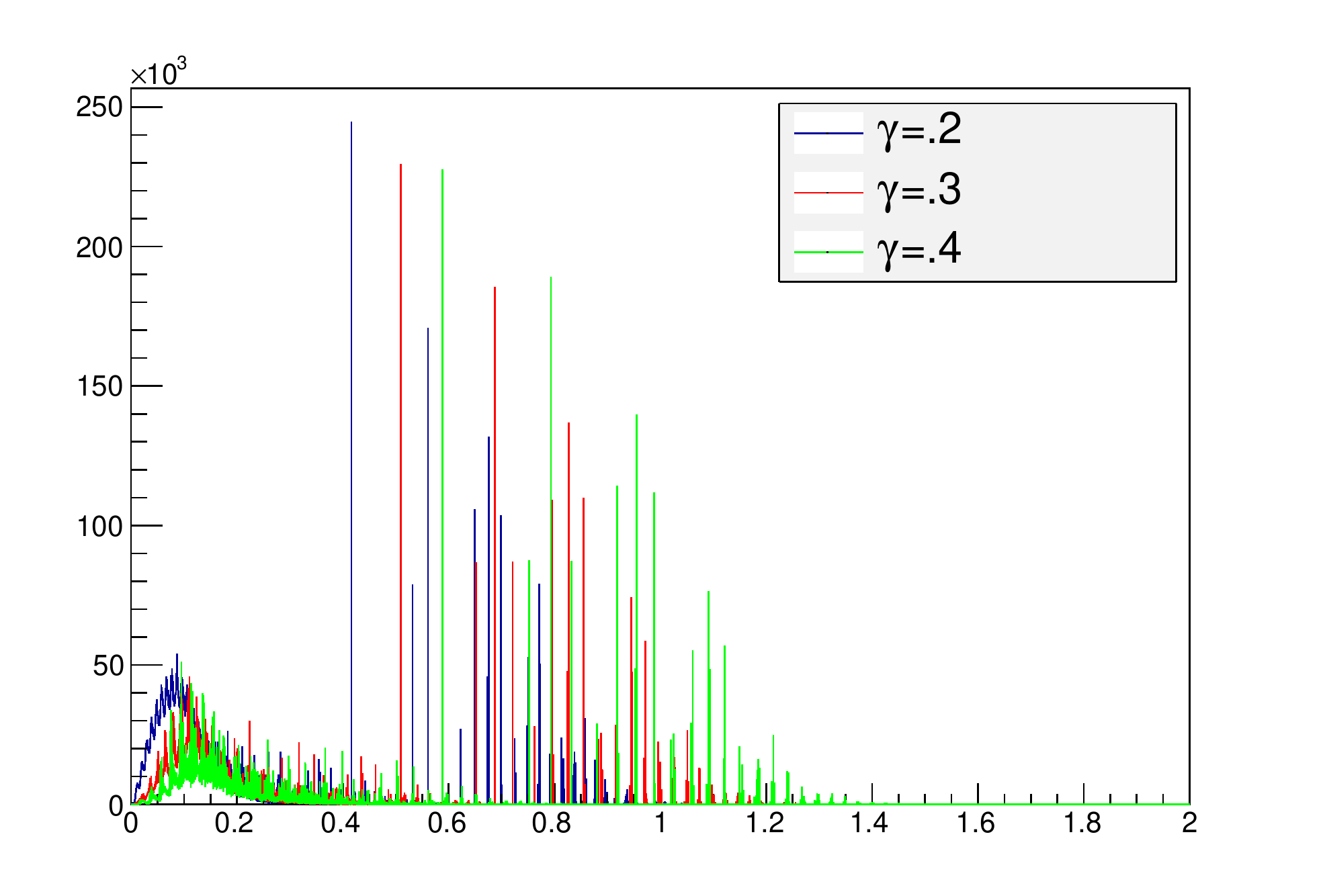}
\end{array}
\)}
\label{oldy}
\caption{$\gamma$ dependence of the integrated spectrum in the standard LQG model (linear scale).} \label{old1}
\end{figure}

\begin{figure}[h] \centerline{\hspace{0.5cm} \(
\begin{array}{c}
\includegraphics[width=9cm]{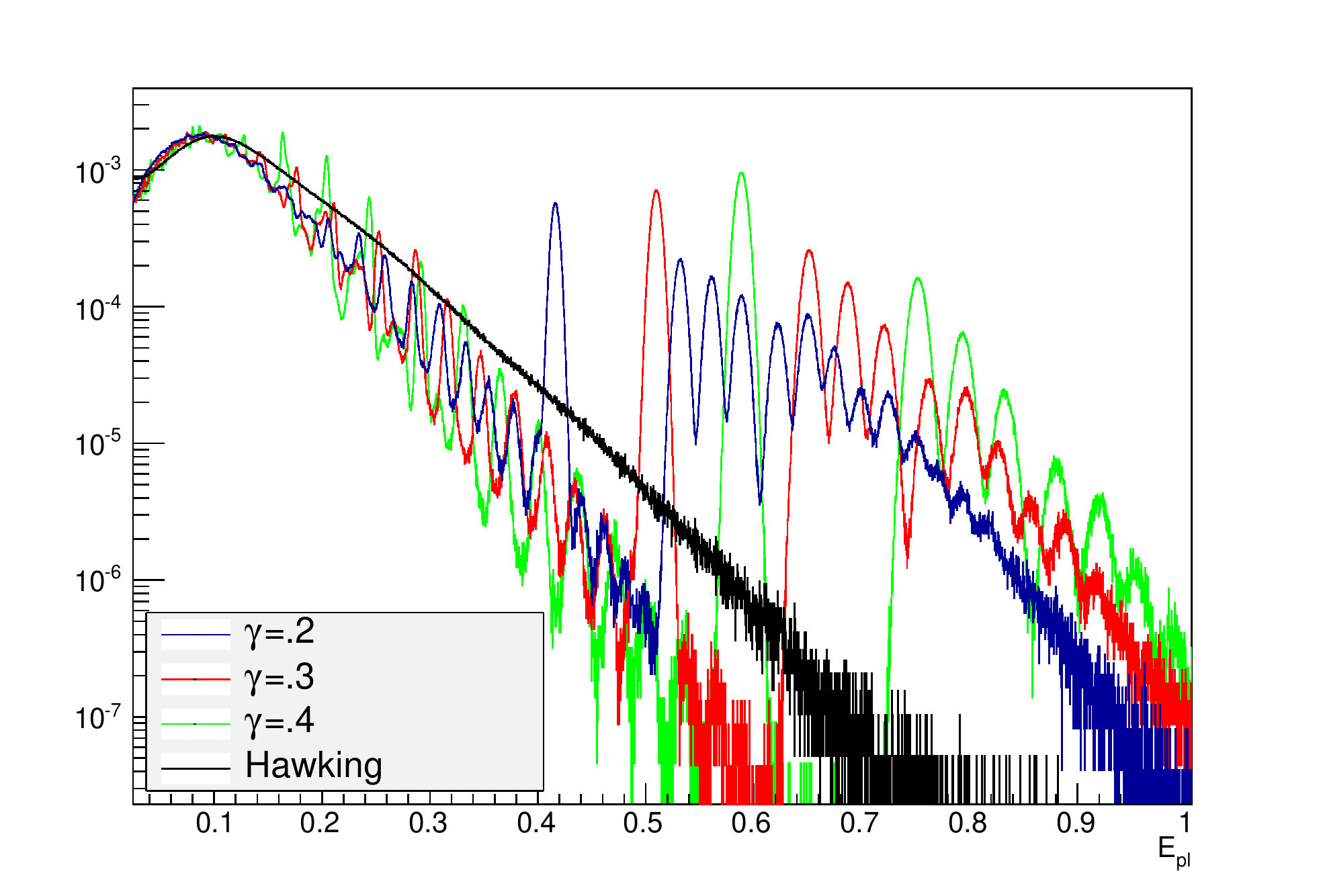}
\end{array}
\)}
\caption{$\gamma$ dependence of the integrated spectrum in the holographic model with a simulated error of energy detection: a particle of energy $E$ has detected energy $E + e$, where $e$ is a normal distribution with variance 
$\sigma = 0.05 E$.} \label{smeared spectrum}
\end{figure}

\begin{figure}[h] \centerline{\hspace{0.5cm} \(
\begin{array}{c}
\includegraphics[width=9cm]{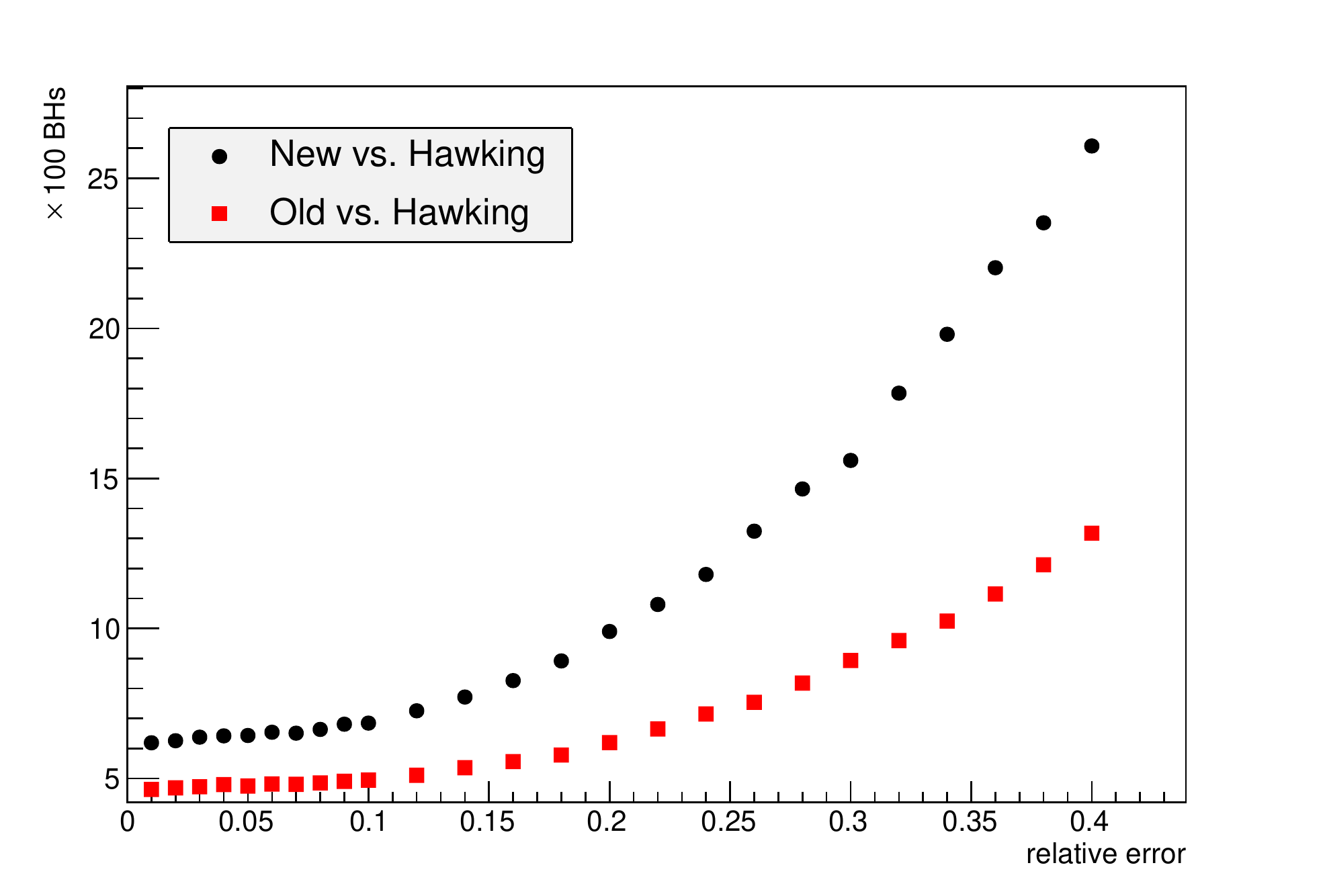}
\end{array}
\)}
\caption{Number of observed events necessary to distinguish the semiclassical Hawking model from LQG models, as a function of the relative uncertainty of the apparatus.} \label{compare}
\end{figure}

A question of physical relevance is how many evaporating black holes  would be necessary to distinguish between the different models studied here and the Hawking semiclassical model. In order to do this in a quite realistic way, we introduced an uncertainty in the energy measurement of 
emitted particles through a random error. With this modification, the spectra in figure \ref{spectrum} gets smeared as shown in figure \ref{smeared spectrum}.  With such smeared spectra is now possible to compute how many detections would be necessary to distinguish the different models. Results are shown in figure \ref{compare}.

In each case, if the number of black holes is high enough and/or the uncertainty small enough, it is possible, through a Kolmogorov-Smirnov test, to statistically distinguish between models. Due to the structure and amplitude of the peaks, it is slightly easier to distinguish between the pure Hawking spectrum and the old LGQ model than between the pure Hawking spectrum and the holographic one.

\section{Conclusions}\label{VI}

The study presented in this article investigates the precise emission from light quantum black holes. Starting with an initial horizon area $A \sim 10^2$ in Planck units, the spectra for different values of $\gamma$ were computed, averaging over many realizations. The continuous background  corresponding to the semi-classical stages is complemented by discrete peaks  associated with the deep quantum regime. We have shown that the Barbero-Immirzi parameter has an important, and sometimes subtle, effect on both parts of the spectrum. Finally we have calculated the number of black holes, for a given experimental uncertainty, required to experimentally distinguish between the holographic model and the semi-classical Hawking spectrum.\\

The analysis presented in this article is more relevant at the conceptual level than at the strictly experimental level. It is more a ``thought experiment", useful to understand subtleties that the analytical investigations do not reveal easily, than a proposal for a real experiment. Is it however conceivable that evaporating black holes can really be used to probe the models ? In the framework of a cosmological production of primordial black holes, this is extremely unlikely (see \cite{Carr:2009jm} for a review). A very wide mass spectrum of primordial black holes is now disfavoured as it would require either a high normalization of the primordial fluctuation power spectrum (to produce a density contrast of order unity) or a blue tilted spectrum, both being excluded by CMB observations. There are many other means for creating primordial black holes with a narrower mass spectrum. In any case, however, the emission spectrum roughly decreases with the energy as $E^{-\alpha}$ with $\alpha$ ranging between around 1  below 100 MeV to around 3 above this scale \cite{Halzen:1991uw,Barrau:2001ev}. Unless a very strong boost phenomenon occurs at late times, it is therefore basically impossible to see the last stages without having first detected, at a much higher level, the semi-classical stage. A more promising avenue would be associated with the production at colliders. This however requires a low Planck scale so that the collision is trans-planckian and creates a black hole. This is not, in principle, incompatible with LQG but relies on strong assumptions --such as extra-dimensions-- that the theory precisely allows to avoid. At this stage, the observational detection is therefore unlikely and simulations are mostly used as {\it gedankenexperiments} to understand better the detailed features of the model.\\

\end{document}